\documentclass[10pt, conference]{IEEEtran}

\usepackage[mathscr]{eucal}
\usepackage[cmex10]{amsmath}
\usepackage{epsfig,epsf,psfrag}
\usepackage{amssymb,amsmath,amsthm,amsfonts,latexsym}
\usepackage{amsmath,graphicx,bm,xcolor,url}
\usepackage{fixltx2e}
\usepackage{array}
\usepackage{verbatim}
\usepackage{bm}
\usepackage{algorithmic, cite}
\usepackage{algorithm}
\usepackage{verbatim}
\usepackage{textcomp}
\usepackage{mathrsfs}
\usepackage{epstopdf}

\catcode`~=11 \def\UrlSpecials{\do\~{\kern -.15em\lower .7ex\hbox{~}\kern .04em}} \catcode`~=13 

\allowdisplaybreaks[4]
 
\newcommand{\nn}{\nonumber}

\newcommand{\calA}{\mathcal{A}}

\newcommand{\calC}{\mathcal{C}}
\newcommand{\calD}{\mathcal{D}}
\newcommand{\calE}{\mathcal{E}}
\newcommand{\calF}{\mathcal{F}}
\newcommand{\calG}{\mathcal{G}}

\newcommand{\calJ}{\mathcal{J}}

\newcommand{\calM}{\mathcal{M}}

\newcommand{\calT}{\mathcal{T}}
\newcommand{\calU}{\mathcal{U}}

\newcommand{\calX}{\mathcal{X}}
\newcommand{\calY}{\mathcal{Y}}
\newcommand{\calZ}{\mathcal{Z}}

\newcommand{\bu}{\mathbf{u}}

\newcommand{\bx}{\mathbf{x}}

\newcommand{\by}{\mathbf{y}}

\newcommand{\rme}{\mathrm{e}}

\newcommand{\rmp}{\mathrm{p}}

\newcommand{\rmr}{\mathrm{r}}

\newcommand{\rms}{\mathrm{s}}

\newcommand{\bbR}{\mathbb{R}}

\newcommand{\scC}{\mathscr{C}}

\newcommand{\scP}{\mathscr{P}}

\newcommand{\scV}{\mathscr{V}}

\DeclareMathAlphabet{\mathbsf}{OT1}{cmss}{bx}{n}
\DeclareMathAlphabet{\mathssf}{OT1}{cmss}{m}{sl}

\DeclareSymbolFont{bsfletters}{OT1}{cmss}{bx}{n}  
\DeclareSymbolFont{ssfletters}{OT1}{cmss}{m}{n}
\DeclareMathSymbol{\bsfGamma}{0}{bsfletters}{'000}
\DeclareMathSymbol{\ssfGamma}{0}{ssfletters}{'000}
\DeclareMathSymbol{\bsfDelta}{0}{bsfletters}{'001}
\DeclareMathSymbol{\ssfDelta}{0}{ssfletters}{'001}
\DeclareMathSymbol{\bsfTheta}{0}{bsfletters}{'002}
\DeclareMathSymbol{\ssfTheta}{0}{ssfletters}{'002}
\DeclareMathSymbol{\bsfLambda}{0}{bsfletters}{'003}
\DeclareMathSymbol{\ssfLambda}{0}{ssfletters}{'003}
\DeclareMathSymbol{\bsfXi}{0}{bsfletters}{'004}
\DeclareMathSymbol{\ssfXi}{0}{ssfletters}{'004}
\DeclareMathSymbol{\bsfPi}{0}{bsfletters}{'005}
\DeclareMathSymbol{\ssfPi}{0}{ssfletters}{'005}
\DeclareMathSymbol{\bsfSigma}{0}{bsfletters}{'006}
\DeclareMathSymbol{\ssfSigma}{0}{ssfletters}{'006}
\DeclareMathSymbol{\bsfUpsilon}{0}{bsfletters}{'007}
\DeclareMathSymbol{\ssfUpsilon}{0}{ssfletters}{'007}
\DeclareMathSymbol{\bsfPhi}{0}{bsfletters}{'010}
\DeclareMathSymbol{\ssfPhi}{0}{ssfletters}{'010}
\DeclareMathSymbol{\bsfPsi}{0}{bsfletters}{'011}
\DeclareMathSymbol{\ssfPsi}{0}{ssfletters}{'011}
\DeclareMathSymbol{\bsfOmega}{0}{bsfletters}{'012}
\DeclareMathSymbol{\ssfOmega}{0}{ssfletters}{'012}

\newcommand{\hate}{\hat{e}}

\newcommand{\hatI}{\hat{I}}

\newcommand{\hatm}{\hat{m}}

\newcommand{\tilm}{\tilde{m}}

\newcommand{\tilR}{\tilde{R}}

\newcommand{\hatV}{\hat{V}}

\newcommand{\ceil}[1]{\lceil{#1}\rceil}

\newcommand{\dotleq}{\stackrel{.}{\leq}}

\newcommand{\dotgeq}{\stackrel{.}{\geq}}

\newcommand{\bone}{\mathbf{1}}

\newtheorem{theorem}{Theorem}

\newcommand{\qednew}{\nobreak \ifvmode \relax \else
      \ifdim\lastskip<1.5em \hskip-\lastskip
      \hskip1.5em plus0em minus0.5em \fi \nobreak
      \vrule height0.75em width0.5em depth0.25em\fi}

\title{Error and Erasure Exponents for the   Broadcast Channel with Degraded Message Sets}
\IEEEoverridecommandlockouts
\author{  
 \IEEEauthorblockN{Vincent Y. F. Tan}
  \IEEEauthorblockA{Department of Electrical and Computer Engineering,\\ Department of Mathematics, \\
    National University of Singapore\\
    Email: \url{vtan@nus.edu.sg} }  
}

\begin{document}
\maketitle
\begin{abstract}   
Error and erasure exponents for the broadcast channel with degraded message sets are analyzed. The  focus of our error probability  analysis is   on the main receiver where, nominally, both messages are to be  decoded. A   two-step decoding algorithm is proposed and analyzed. This receiver first attempts to decode both messages, failing which, it attempts to decode only the message representing the coarser information, i.e., the cloud center.  This algorithm reflects the intuition that we should decode  both messages only if we have  confidence in the estimates; otherwise one should only decode the coarser information.  The resulting error and erasure exponents, derived using the method of types, are expressed in terms of a penalized form of the modified random coding error exponent. 
\end{abstract}


\section{Introduction}\label{sec:intro}
In 1968, Forney \cite{Forney68} derived exponential error bounds for decoding with an erasure option. In this seminal paper, Forney used a generalization of the Neyman-Pearson lemma to derive an optimum decoding rule for point-to-point channel coding, where the decoder is allowed to output an erasure symbol should it not be sufficiently confident to output a message. Based on this rule, Forney used Gallager-style bounding techniques to derive  exponents  for   the undetected and total  (undetected plus erasure)  error probabilities. 

This work led to many follow-up works. We only mention a subset of the literature here. We mainly build on the exposition in Csisz\'ar and K\"orner \cite[Thm.~10.11]{Csi97} in which universally attainable erasure and error exponents were derived. Telatar~\cite{telatar_thesis} also derived and analyzed a erasure decoding rule  based on a general decoding metric. Moulin~\cite{Moulin09} considered a Neyman-Pearson formulation for universal erasure decoding.  Merhav \cite{merhav08} used the   type-class enumerator method to analyze the Forney decoding rule and showed that the derived exponents are at least as good as those Forney derived. This was subsequently sharpened   by Somekh-Baruch and Merhav~\cite{somekh11} who derived the {\em exact} random coding exponents. Sabbag and Merhav~\cite{sabbag}   analyzed the error and erasure exponents for channels with noncausal state information (Gel'fand-Pinsker coding).

However, no generalization of the study of erasure exponents to multi-user systems with multiple messages has been published.\footnote{Moulin mentioned in~\cite[Sec.~VIII]{Moulin09} that the analysis contained therein ``has been extended to compound MACs'' but this extension is   unpublished.   } In this paper, we study the broadcast channel with degraded message sets, also known as the asymmetric broadcast channel (ABC). For this channel, the main receiver desires to decode two messages $M_1$ and $M_2$ while the secondary receiver only desires to decode the private message $M_2$. The capacity region, derived by K\"orner and Marton  \cite{kor77} is  
\begin{align}
\scC\!=\!\bigcup_{P_{UX}}\left\{\! (R_1,R_2)\! \in\!\bbR_+^2\!:\! \begin{array}{rl}
R_1   &\hspace{-.1in}\le I(X\wedge Y|U)\\
R_2 &\hspace{-.1in}\le I(U\wedge Z)\\
R_1+R_2 &\hspace{-.1in}\le I(X\wedge Y) 
\end{array}\!\!\!\right\}  . \label{eqn:rr1}
\end{align}
Error exponents (without erasures) were   derived by K\"orner and Sgarro \cite{Kor80b} and improved    by Kaspi and Merhav \cite{kaspi10}.  We go beyond these analyses to derive   erasure and error exponents for the ABC. The resulting exponents involve a {\em penalized} form  of the modified   random coding error exponent derived in~\cite[Ch.\ 10]{Csi97} and reflects the superposition coding scheme \cite{cover72} used to achieve the region in \eqref{eqn:rr1}.
\section{Preliminaries and System Model}\label{sec:prelims}

We adopt the notation from~\cite{Csi97}. Random variables  (e.g., $X$)   and their realizations (e.g., $x$)   are in upper- and  lower-case respectively. All random variables take values on finite sets,  denoted in calligraphic font (e.g., $\calX$). For a sequence $\bx=(x_1,\ldots, x_n)\in\calX^n$, its {\em type} is the distribution $P_\bx(a)=\frac{1}{n}\sum_{i=1}^n\bone\{x_i=a \},a\in\calX$. The set of types with denominator $n$ supported on  alphabet $\calX$ is denoted as $\scP_n(\calX)$. The {\em type class} of $P$ is  denoted as $\calT_P$.  For     $\bx \in\calT_P$,  the set of sequences $\by \in\calY^n$ such that $(\bx, \by)$ has joint type $P \times V$ is the {\em $V$-shell} $\calT_V(\bx)$.  Let $\scV_n(\calY;P)$  be the family of stochastic matrices $V\!:\!\calX\!\to\!\calY$ for which the $V$-shell of a sequence of type $P\in\scP_n(\calX)$ is  not empty. Information-theoretic quantities are denoted in the usual way. For example  $I(P,V)$ and $ I_{P\times V}(X\wedge Y) $ denote the mutual information  where these expressions indicate that the joint  distribution of $(X,Y)$ is      $P\times V$. In addition, $\hatI(\bx\wedge \by)$ is the empirical mutual information of   $(\bx, \by)$, i.e., if  $\bx   \in \calT_{P  }$ and $ \by\in\calT_V(\bx)$, then,  $\hatI(\bx\wedge \by):=I(P,V)$.  We use $a_n\doteq b_n$ to mean {\em equality to first-order in the exponent}, i.e., $\frac{1}{n}\log\frac{a_n}{b_n}\to 0$;   exponential inequalities $\dotleq$ and $\dotgeq$ are defined similarly. Finally, $|a|^+ :=\max\{a, 0\}$ and $[a]:=\{1,\ldots, \ceil{ a}\}$  for any $a\in\bbR$. 

A {\em discrete memoryless broadcast channel} $W:\calX\to\calY\times\calZ$ is a stochastic map from a finite input alphabet $\calX$ to the Cartesian product of two finite output alphabets $\calY$ and $\calZ$. An {\em  $(n,R_1, R_2)$-code} is a tuple of maps $f: [2^{nR_1}]\times [2^{nR_2}]\to \calX^n$ and $\varphi_1 : \calY^n\to ([2^{nR_1}]\cup \{\rme_1\})\times ( [2^{nR_2}]\cup \{\rme_2\})$ and $\varphi_2 :\calZ^n \to  [2^{nR_2}]\cup\{\rme_2\}$, where $\rme_j$ is the erasure symbol for message $M_j , j=1,2$ which is uniform on the {\em message set} $\calM_j := [2^{nR_j}]$. Let $W_\calY$ and $W_\calZ$ be the marginals of~$W$.  

We may define error probabilities for both terminals $\calY$ and $\calZ$. However, we will focus exclusively on terminal $\calY$ as the analysis is more interesting and non-standard. We are concerned with four different error probabilities at terminal $\calY$. Let $\calD_{m_1}, m_1 \in \calM_1:=[2^{nR_1}]$ and $\calD_{m_2}, m_2\in  \calM_2:=[2^{nR_2}]$  be the disjoint decoding regions associated to messages $m_1$ and $m_2$ respectively. This means that $\calD_{m_1} := \cup_{m_2\in\calM_2}\{\by: \varphi_1(\by)=(m_1, m_2)\}$ and similarly for $\calD_{m_2}$. Note that because we allow erasures $\calY^n\setminus \cup_{m_1\in\calM_1} \calD_{m_1}$ need not be an empty set. Define for message $j = 1,2$, the   conditional total (undetected plus erasure) and undetected error probabilities at terminal $\calY$ 
\begin{align}
\xi_j(m_1,m_2) &:=W_\calY^n\big(   \calD_{m_j}^c\, \big|\, \bx(m_1, m_2) \big)  \\
\hat{\xi}_j(m_1,m_2) &:=W_\calY^n\Big(    \bigcup_{ \tilm_j \in \calM_j\setminus \{m_j\}} \calD_{\tilm_j} \, \big|\, \bx(m_1, m_2) \Big)  .
\end{align}
Then we may define the   average total and undetected error probabilities for message  $j$ at terminal $\calY$  as follows:
\begin{align}
e_j& :=\frac{1}{|\calM_1||\calM_2|}\sum_{(m_1,m_2)  \in  \calM_1\times\calM_2} \xi_j(m_1, m_2)\\
\hate_j & :=\frac{1}{|\calM_1||\calM_2|}\sum_{(m_1,m_2) \in \calM_1\times\calM_2}\hat{\xi}_j(m_1, m_2).
\end{align}
The objective of this paper is to find exponential upper bounds for $(e_1, \hate_1,e_2,\hate_2)$, all of which depend on the blocklength $n$.

\section{Decoding Strategy} \label{sec:strategy}
In this section, we detail the decoding strategy at terminal $\calY$. The decoding strategy and subsequent analysis  for terminal $\calZ$ is standard and follows from Csisz\'ar and K\"orner's exposition of decoding with the erasure option~\cite[Thm.~10.11]{Csi97}. 

Assume there is a codebook $\calC$ consisting  {\em cloud centers} $\bu(m_2) \in\calU^n, m_2\in\calM_2$ ($\calU$ is a finite set) and for each $m_2$, a set of {\em satellite codewords} $\bx(m_1, m_2)  \in\calX^n$ indexed by $m_1\in\calM_1$. Fix $\lambda_{12},\lambda_2\ge 1$ and $\tilR_j\ge R_j$ for $j=1,2$. For brevity, let $\tilR_{12}:=\tilR_1+ \tilR_2$ and $R_{12}:=R_1+R_2$ be the sum rates. The decoding rule is given as follows:

Step 1: Decode to $(\hatm_1, \hatm_2) \in \calM_1 \times \calM_2$ if and only if this is the {\em unique} pair of messages such that 
\begin{align}
&\hatI( \bu(\hatm_2) , \bx(\hatm_1, \hatm_2) \wedge \by)  \nn\\*
 &\quad\ge\tilR_{12}+\lambda_{12}  \big|\hatI( \bu(\tilm_2) , \bx(\tilm_1, \tilm_2) \wedge \by) - R_{12} \big|^+ \label{eqn:Step1}
\end{align}
for all $(\tilm_1, \tilm_2)\ne (\hatm_1, \hatm_2)$.  If we cannot find a unique pair of messages satisfying~\eqref{eqn:Step1}, go to Step 2.

Step 2: Declare the first message to be an erasure $\rme_1$ and declare the second message to be $\hatm_2 \in \calM_2$ if and only if it is the {\em unique} message such that  
\begin{equation}
\hatI( \bu(\hatm_2)   \wedge \by) \ge \tilR_{ 2} +\lambda_{ 2}  \big|\hatI( \bu(\tilm_2)  \wedge \by) - R_{2} \big|^+ \label{eqn:Step2}
\end{equation}
for all $\tilm_2 \ne  \hatm_2$.
If we cannot find a unique message satisfying \eqref{eqn:Step2}, declare the second message to be an erasure $\rme_2$ as well.

The intuition behind this   two-step algorithm is as follows: In Step 1, we are ambitious. We try to  decode both messages $M_1$ and $M_2$ using the   rule   in~\eqref{eqn:Step1}. This   rule is a generalization of that for the single-user case in   \cite[Thm.~10.11]{Csi97}. If decoding {\em fails} (i.e., no unique message pair   satisfies \eqref{eqn:Step1}), perhaps due to the stringent choices of $\tilR_{1}$, $\tilR_{2}$  and $\lambda_{12}$, then we act  conservatively. Given $\by$,  we {\em at least} want to decode the cloud center represented by  $M_2$, while we are content with declaring an erasure for $M_1$. If Step 2 in \eqref{eqn:Step2} also fails,   we have no choice but to erase both messages. Note that the decoding rules in~\eqref{eqn:Step1}--\eqref{eqn:Step2} are unambiguous because $\lambda_{12},\lambda_2\ge 1$~\cite[App.~I]{sabbag}.

\section{Main Result and Interpretation}\label{sec:main_res}
\subsection{Preliminary Definitions}
Before we present the main result, we define a few relevant quantities. First, we fix a joint distribution $P_{UX}\in\scP (\calU\times\calX)$. Next fix   conditional distributions $V :\calU\times\calX\to\calY$ and $\hatV:\calU\to\calY$. Then we may define
\begin{align}
J_V(R_1, R_2)  &:= |I_V(UX\wedge Y)-R_{12}|^+ \label{eqn:JV}\\
J_{\hatV}( R_2)  &:= |I_{\hatV}(U \wedge Y)-  R_2 |^+ \label{eqn:JVhat} .
\end{align}
Note that $I_{V}(UX\wedge Y)$ is the mutual information of $UX$ and $Y$ where the joint distribution of $UXY$ is $P_{UX}\times V$ but $P_{UX}$, being fixed throughout, is suppressed in the notations in \eqref{eqn:JV} and \eqref{eqn:JVhat}. 
We define the {\em marginal and joint modified random coding error exponents for the ABC} as
\begin{align}
E_{\rmr,\lambda}(\tilR_2 ) &:= \min_{\hatV} D(\hatV\| W_{Y|U} | P_{U }) + \lambda J_{\hatV}(\tilR_2) \label{eqn:rcee_mar} \\
E_{\rmr,\lambda}(\tilR_1, \tilR_2 ) &:= \min_V D(V\| W | P_{UX}) + \lambda J_V(\tilR_1, \tilR_2) . \label{eqn:rcee_jt}
\end{align}
Here, we use the notation $W_{Y|U} $ and $W$ to mean the channels $W_{Y|U}(y|u) := \sum_x W_\calY(y|x)P_{X|U}(x|u)$ and  $W(y|u,x):=W_\calY(y|x)$ for every $u\in\calU$. Also note that the exponents in \eqref{eqn:rcee_mar} and \eqref{eqn:rcee_jt} depend on $P_{UX}$ but this dependence is suppressed.  Furthermore, we define the {\em penalized modified random coding error exponent for the ABC} as 
\begin{align}
&E_{\rmr,\lambda}^{-}(\tilR_1, \tilR_2 , R_2 )\nn\\* 
& \quad :=\min_V D(V\| W | P_{UX})   + \lambda J_V(\tilR_1, \tilR_2)- J_V(R_2) . \label{eqn:pen_rcee}
\end{align}
The penalization comes from the fact that we are subtracting the non-negative quantity $J_V(R_2)$ in the optimization above. 
Define the {\em sphere packing exponent for the ABC} as 
\begin{equation}
E_{\mathrm{sp}}(R):= \min_{V: I_V(UX\wedge Y)\le R}  D(V\| W | P_{UX}) . \label{eqn:sp_exp}
\end{equation}
Finally, for $\tilR_{2}$ and $R_2$, we define the {\em difference in rates} $\Delta_2 =\Delta_2(\tilR_2,R_2):= \tilR_2-R_2$ and similarly,  $\Delta_{12}=\Delta_{12}(\tilR_{1},R_{1}, \tilR_{2},R_{2}):=\tilR_{12}-R_{12}$.
\subsection{Main Result} 
With these preparations, we can now state our main result.
\begin{theorem} \label{thm:exp}
There exists a sequence of $(n,R_1, R_2)$-codes for the ABC such that for any choice of   $\tilR_{1}$, $\tilR_{2}$, $\lambda_{12}$, $\lambda_2$ and $P_{UX}$, we have 
\begin{align}
\liminf_{n\to\infty} \frac{1}{n}\log\frac{1}{e_1} &\ge E_{\rmr,\frac{1}{\lambda_{12}}}^{-}(\tilR_1,\tilR_2, R_2) \label{eqn:e1_res} \\
\liminf_{n\to\infty} \frac{1}{n}\log\frac{1}{\hate_1} &\ge E_{\rmr, \lambda_{12}}^{-}(R_1,R_2, R_2)+\Delta_{12}  \label{eqn:e1hat_res} \\
\liminf_{n\to\infty} \frac{1}{n}\log\frac{1}{e_2} &\ge  \min\Big\{ E_{\rmr,\lambda_{12}}(R_1,R_2) +\Delta_{12},  \nn\\*
&\hspace{-.55in}\max\big\{E_{\rmr, \frac{1}{\lambda_{12}}}^{-}  ( \tilR_1,\tilR_2 ,R_2),  E_{\rmr, \frac{1}{\lambda_2}}(\tilR_2)  \big\}\Big\} \label{eqn:e2_res}\\
\liminf_{n\to\infty} \frac{1}{n}\log\frac{1}{\hate_2} &\ge \min\Big\{  E_{\rmr , \lambda_{12}}(R_1,R_2 ) + \Delta_{12},\nn\\*
&\hspace{-.55in}\max\big\{E_{\rmr, \frac{1}{\lambda_{12}}}^{-}  ( \tilR_1,\tilR_2 , R_2),   E_{\rmr,\lambda_2}(R_2)+\Delta_2 \big\}
\Big\}
 \label{eqn:e2hat_res}
\end{align}
\end{theorem}

\subsection{Interpretation and Comments}
We now comment on the forms  of the exponents. 

First, observe from \eqref{eqn:e1_res} and \eqref{eqn:e1hat_res} involve the penalized modified random coding error exponent for the ABC $E_{\rmr, \lambda }^{-}$, defined in \eqref{eqn:pen_rcee}. The penalization of the term $J_V(R_2)$ is required because of there are various ways an undetected error and erasure can occur for message 1 according to Step 1 of the decoding rule. More specifically, this penalization results from the kinds of errors that may result from superposition coding~\cite{cover72}: (i) the cloud center may be decoded incorrectly, or (ii) the cloud center is decoded correctly but the satellite is decoded incorrectly. The latter event leads to the term $J_V(R_2)$.

Second, since $\tilR_j\ge R_j$ and $\lambda_{12},\lambda_2\ge 1$, the undetected exponents are at least as large as  the error exponents. This is natural since undetected errors are much more undesirable compared to erasures. We can simply re-transmit if there is an erasure. This is   known as Automatic Repeat reQuest (ARQ).

Third, observe from \eqref{eqn:e2_res} and \eqref{eqn:e2hat_res} that the exponents associated to message 2 involve several terms. Since the algorithm we proposed in Sec.~\ref{sec:strategy}  involves two distinct steps,  undetected errors and erasures for message 2 occur in several different ways  so the event with the worst exponent dominates. 

Finally, we mention that unlike in Csisz\'ar and K\"orner~\cite[Ch.\ 10]{Csi97} and K\"orner and Sgarro~\cite{Kor80b}, we do not  (explicitly) use the packing lemma (or its generalizations) in our proof. We compute bounds on the error and erasure probabilities averaged over a random constant joint-composition code directly.
 
 \section{Proof of Main Result}\label{sec:prf}
The proof is split into four subsections, each detailing the calculations leading to the exponents presented in Theorem~\ref{thm:exp}. 

Given joint type $P_{UX}\in\scP_n(\calU\times\calX)$, we consider the ensemble of codes that are randomly generated as follows: First for each $m_2\in [2^{nR_2}]$, a cloud center $\bu(m_2)$ is drawn uniformly at random from the type class $\calT_{P_{U}}\subset\calU^n$. Subsequently for each $m_2$, a cloud of codewords called satellite codewords $\bx(m_1, m_2), m_1 \in [2^{nR_1}]$ is drawn uniformly at random from the $P_{X|U}$-shell $\calT_{P_{X|U}}(\bu(m_2))\subset\calX^n$.  We analyze the error probabilities averaged over this random codebook construction and we assume, without loss of generality, that the transmitted messages are $(M_1, M_2)=(1,1)$. 

\subsection{Undetected Error Probability for Message 1 at Terminal $\calY$} \label{sec:und1}
An undetected error for message 1 occurs if and only if Step 1 succeeds (i.e., outputs a message pair instead of going on to Step 2) but the declared message pair $(\hatm_1, \hatm_2)$ is such that the first component is some natural number other than 1 (second component is arbitrary). In other  words, we have 
\begin{align}
\hate_1  = \Pr\bigg(\bigcup_{\tilm_1 \in\calM_1\setminus\{1\}, \tilm_2\in\calM_2} \calE(\tilm_1,\tilm_2) \bigg)
\end{align} 
where the event $\calE(\tilm_1,\tilm_2) $ is defined as 
\begin{align}
&\calE(\tilm_1,\tilm_2) :=\big\{\hatI( U^n(\tilm_2) , X^n(\tilm_1, \tilm_2) \wedge Y^n) \nn\\*
&\ge \tilR_{12} +\lambda_{12}   \big|\hatI( U^n(1) , X^n(1, 1) \wedge Y^n) - R_{12} \big|^+ \big\} \label{eqn:Em1m2}
\end{align}
To analyze $\hate_1$, we first condition on $(U^n(1), X^n(1,1), Y^n)$ having various joint types, i.e.,
\begin{align}
&\hate_1 := \sum_{V_{UXY}} \sum_{(\bu,\bx,\by) \in \calT_{V_{UXY}}} P_{U^n X^n}(\bu,\bx)W_{\calY}^n(\by|\bx) \nn\\*
&\times \Pr\bigg(\bigcup_{\tilm_1 \in\calM_1\setminus\{1\}, \tilm_2\in\calM_2} \calE(\tilm_1,\tilm_2) \, \bigg|\, \bu,\bx,\by \bigg)\label{eqn:cond_type} 
\end{align} 
The first sum over $V_{UXY}$ is in fact over all joint types in $\scP_n(\calU\times\calX \times \calY)$ for which the $(\calU\times\calX)$-marginal is $P_{UX}$. 
The conditioning in the probability in \eqref{eqn:cond_type} is on the event $\{ U^n(1) = \bu,X^n(1,1)=\bx, Y^n=\by\}$ but we shorten this to $\{ \bu,\bx,\by \}$ for brevity. 

Now we distinguish between    two cases: Case (i) $\tilm_1\ne 1, \tilm_2\ne 1$,  and  Case (ii) $\tilm_1\ne 1 ,\tilm_2=1$.  For Case (i), there are  $\lceil 2^{nR_{12}}\rceil-1$ such events and by symmetry we may analyze  
\begin{align}
&\Pr\big( \calE(2, 2) \, \big|\, \bu,\bx,\by \big)\nn\\*
&=\Pr\big(\hatI( U^n(2) , X^n(2, 2) \wedge Y^n) \ge t\, \big|\, \bu,\bx,\by  \big)  \label{eqn:calE22}
\end{align}
where given $(\bu,\bx,\by )$, the parameter 
\begin{align}
t 
 = \tilR_{12} +\lambda_{12}   \left|I_V( UX \wedge Y ) - R_{12} \right|^+ \label{eqn:def_t}
\end{align}
  is fixed. We suppress the dependence of $t$ on  $(\bu,\bx,\by )$. Now we bound the   probability  in \eqref{eqn:calE22} as follows:
\begin{align}
&\Pr\big( \calE(2, 2) \, \big|\, \bu,\bx,\by \big)  \nn\\*
&= \sum_{\substack{\hatV_{UX|Y} \in \scV_n(\calU\times\calX;P_{\by}):\\ P_{\by}\hatV_{UX|Y}=P_{UX}, I(P_{\by},\hatV_{UX|Y})\ge t }}P_{U^nX^n} (\calT_{\hatV_{UX|Y}}(\by)) \\
 &\doteq\sum_{\substack{\hatV_{UX|Y} \in \scV_n(\calU\times\calX; P_{\by}):\\ P_{\by}\hatV_{UX|Y}=P_{UX}, I(P_{\by},\hatV_{UX|Y})\ge t }}\exp(-n I(P_{\by},\hatV_{UX|Y})) \label{eqn:E22_prev}\\
 &\doteq \exp(-nt)\label{eqn:E22}
\end{align}
where \eqref{eqn:E22_prev} follows from a standard method of types calculation.  See, for example, \cite[Lem~10.1]{Csi97} or \cite[Appendix]{TanRelay}.
Next consider Case (ii). In this case there are at most $2^{n R_1 }$ such events which indicates that the cloud center is  decoded correctly but the satellite codeword is not. The conditional probability of a generic event in this case $\calE(2, 1)$ can be bounded as follows:
\begin{align}
&\Pr\big( \calE(2, 1) \, \big|\, \bu,\bx,\by \big) \nn\\*
 &= \sum_{\substack{\hatV_{X|UY} \in \scV_n(\calX;P_{\bu,\by}):\\ P_{\by|\bu }\hatV_{ X|UY}=P_{ X|U}, \\ I(P_{\bu,\by},\hatV_{X|UY}  )\ge t }}P_{X^n|U^n} (\calT_{\hatV_{ X|UY}}(\bu,\by) | \bu) \\
&\doteq \sum_{\substack{\hatV_{X|UY} \in \scV_n(\calX;P_{\bu,\by}):\\ P_{\by|\bu}\hatV_{ X|UY}=P_{ X|U}, \\ I(P_{\by|\bu},\hatV_{X|UY} |P_{\bu})+ I(P_{\bu}, P_{\by|\bu})\ge t }}\exp(-n I( P_{\by|\bu}, \hatV_{X|UY}|P_{\bu}))\\
&\doteq\exp(-n [t- I_V(U\wedge Y) ]) \label{eqn:E21}
\end{align}
where in the last step, we recall that the joint type of $(U^n(1), X^n(1,1), Y^n)$ is $V_{UXY}$ so we have the equality $I_V(U\wedge Y) = I(P_{\bu}, P_{\by|\bu})$. Note that the exponent here is $t-I_V(U\wedge Y)$, which is different from the exponent resulting from the  calculation leading to \eqref{eqn:E22}, i.e., simply $t$. 
Putting the bounds in \eqref{eqn:E22} and \eqref{eqn:E21} together and applying the union bound to the probability in \eqref{eqn:cond_type}, we obtain
\begin{align}
&\Pr\bigg(\bigcup_{\tilm_1\ne 1, \tilm_2} \calE(\tilm_1,\tilm_2) \, \bigg|\, \bu,\bx,\by \bigg)\nn\\*
&  \le 2^{n R_{12}} \Pr\big( \calE(2, 2) \, \big|\, \bu,\bx,\by \big) + 2^{n  R_1}  \Pr\big( \calE(2, 1) \, \big|\, \bu,\bx,\by \big) \\
&\doteq \exp( - n \min\{  t-R_{12} , t-R_1-  I_V(U\wedge Y) \})\\
&= \exp \big( - n \big( t- R_{12}-    \big| I_V(U\wedge Y ) -R_2 \big|^+ \big)  \big) \label{eqn:plus}  . 
\end{align}
Since $P_{U^nX^n}(\bu,\bx)\dotleq P_{UX}^n (\bu,\bx)$ for every $(\bu,\bx)$,  
\begin{equation}
 \sum_{(\bu,\bx,\by) \in \calT_{V_{UXY}}} \!\!\!\!\! P_{U^n X^n}(\bu,\bx)W_{\calY}^n(\by|\bx)\!\dotleq\!  \exp(-n D(V \| W |P_{UX})). \label{eqn:prob_types}
\end{equation}
Since there are only polynomially many joint types~\cite[Ch.~2]{Csi97}, uniting~\eqref{eqn:def_t}, \eqref{eqn:plus} and \eqref{eqn:prob_types}, we conclude that  the undetected error exponent for decoding message 1 is
\begin{align}
&\liminf_{n\to\infty} \frac{1}{n}\log \frac{1}{\hate_1 }  \nn\\*
&\ge \min_{V }  D(V  \| W |P_{UX})+\lambda_{12}   \left|I_V( UX \wedge Y )-R_{12}  \right|^+ \nn\\* 
&\qquad -    \left| I_V(U\wedge Y) -R_2 \right|^+ +   \Delta_{12}    \\
&= E_{\rmr, \lambda_{12}}^{-}( R_1,R_2,R_2)+ \Delta_{12} ,
\end{align}
where the last line arises from the definition  of the penalized modified random coding exponent in \eqref{eqn:pen_rcee}. This gives \eqref{eqn:e1hat_res}.

\subsection{Total Error Probability for Message 1 at Terminal $\calY$} \label{sec:total1}
An error for message 1 occurs under two possible conditions: (i) Step 1 succeeds, in which case we have an undetected error since $m_1$ is declared to be some natural number not equal  to 1; (ii) Step 1 fails,  in which case $m_1$ is erased.  We already analyzed Case (i) in Sec.~\ref{sec:und1} and because $\tilR_j\ge R_j, j = 1,2$ and $\lambda_{12}\ge 1$, this case will not dominate, i.e., its exponent will be larger than that for Case (ii). Hence, we focus on Case (ii), i.e., there is not unique message pair that satisfies \eqref{eqn:Step1}. In particular, message pair $(1,1)$ does  {\em not} satisfy \eqref{eqn:Step1}.  Thus, 
\begin{align}
 e_1 \doteq \Pr\bigg(\bigcup_{(\tilm_1 ,\tilm_2)\in (\calM_1\setminus\{1\}) \times (\calM_2\setminus\{1\})} \calJ(\tilm_1,\tilm_2) \bigg)
\end{align} 
where the event $\calJ(\tilm_1,\tilm_2)$ is defined as follows:
\begin{align}
&\calJ(\tilm_1,\tilm_2):=\big\{\hatI( U^n(1) , X^n(1, 1) \wedge Y^n) \nn\\*
& \!\le \!\tilR_{12} \!+\!\lambda_{12}   \big|\hatI( U^n(\tilm_2) , X^n(\tilm_1, \tilm_2)\! \wedge\! Y^n)\! -\! R_{12} \big|^+ \big\} .
\end{align}
Now similarly to \eqref{eqn:cond_type}, we again partition  into various joint types, i.e., 
\begin{align}
 e_1   &\doteq \sum_{V_{UXY}} \sum_{(\bu,\bx,\by) \in \calT_{V_{UXY}}} P_{U^n X^n}(\bu,\bx)W_{\calY}^n(\by|\bx)  \nn\\*
 &\hspace{-.2in}\times \Pr\bigg(\bigcup_{(\tilm_1 ,\tilm_2)\in (\calM_1\setminus\{1\}) \times (\calM_2\setminus\{1\})} \!\!\!\!\!\!\!\!\calJ(\tilm_1,\tilm_2) \,\bigg|\,  \bu,\bx,\by\bigg)\\
&=:  e_{1,\calA} + e_{1,\calA^c} . \label{eqn:par_A}
\end{align} 
In \eqref{eqn:par_A}, we split the analysis into two parts by partitioning the joint types $V_{UXY}=P_{UX}\times V \in \scP_n(\calU\times\calX\times\calY)$ into two classes:   $\calA:=\{V_{UXY}:  I(P_{UX}, V )\le \tilR_{12}\}$ and   $\calA^c:=\{V_{UXY}: I(P_{UX}, V ) > \tilR_{12}\}$. The first class   results in a  sphere packing-like bound. More precisely,  by Sanov's theorem \cite[Prob.~2.12]{Csi97},
\begin{align}
e_{1,\calA} &= \sum_{V_{UXY} \in \calA} \sum_{(\bu,\bx,\by) \in \calT_{V_{UXY}}} P_{U^n X^n}(\bu,\bx)W_{\calY}^n(\by|\bx) \\
&\doteq \exp(-n E_{\rms\rmp}(\tilR_{12}  )). \label{eqn:sanov}
\end{align}
In the last line, we employed the definitions of $\calA$ and that of the sphere packing exponent in \eqref{eqn:sp_exp}.

Now we deal with the other joint types, i.e., those in $\calA^c$.  Again, we partition the analysis into three cases: Case (i) $\tilm_1\ne 1,\tilm_2\ne 1$;   Case (ii) $\tilm_1 = 1,\tilm_2\ne 1$; Case (iii) $\tilm_1 = 1,\tilm_2\ne 1$.

For Case (i), there are a total of $2^{nR_{12}}$ events with identical probability. A generic such probability is 
\begin{align}
&\Pr\big( \calJ(2, 2) \, \big|\, \bu,\bx,\by \big)\nn\\*
&\le \Pr \big(  \hatI( U^n(2) , X^n(2, 2) \wedge Y^n )\ge s \, \big|\, \bu,\bx,\by \big)\label{eqn:int_s}
\end{align}
where now  because the joint type $V_{UXY}$ is in $\calA^c$, $\hatI(U^n(1) , X^n(1, 1) \wedge Y^n ) > \tilR_{12}$ (i.e., the $|\cdot|^+$ is inactive), the parameter $s$  in \eqref{eqn:int_s} is defined as 
\begin{equation}
s:=\frac{1}{\lambda_{12}}  | I_V(U X \wedge Y )-\tilR_{12} |^+ + R_{12}.
\end{equation}
By the same reasoning as the steps leading to \eqref{eqn:E22}, we have 
\begin{equation}
\Pr\big( \calJ(2, 2) \, \big|\, \bu,\bx,\by \big) \dotleq \exp(-ns) .\label{eqn:F22}
\end{equation}

For Case (ii), there are a total of $2^{nR_1}$ events with identical probability.  Similarly to the calculation that led to \eqref{eqn:E21} and \eqref{eqn:F22}, we have that 
\begin{equation}
\Pr\big( \calJ(2, 1) \, \big|\, \bu,\bx,\by \big) \dotleq \exp(-n [s-I_V(U\wedge Y)])\label{eqn:F21}
\end{equation}

For Case (iii),   there are a total of $2^{nR_2}$ events with identical probability.  Note that here the cloud center, represented by $\tilm_2$, is incorrect, so similarly to the calculation that led to \eqref{eqn:F22}, we have that 
\begin{equation}
\Pr\big( \calJ(1, 2) \, \big|\, \bu,\bx,\by \big) \dotleq \exp(-ns) .\label{eqn:F12}
\end{equation}
Case (iii), however, will not dominate since there are only $2^{nR_2}$ events each with probability given by \eqref{eqn:F12}. This is dominated by Case (i) in which there are exponentially many more error events ($2^{nR_{12}}$ to be precise) with the same bound on the error probability. So we may safely omit the contribution by Case (iii). 

Putting the analysis for the joint types $V_{UXY}\in\calA^c$ together and using \eqref{eqn:prob_types}, we obtain that 
\begin{equation}
\liminf_{n\to\infty}\frac{1}{n}\log\frac{1}{e_{1,\calA^c}}\ge E_{\rmr, \frac{1}{\lambda_{12}}}^{-} ( \tilR_1,\tilR_2,R_2)
\end{equation}
By using the fact that $|\cdot|^+\ge 0$,
\begin{equation}
E_{\rmr, \frac{1}{\lambda_{12}}}^{-} ( \tilR_1,\tilR_2,R_2 ) \le E_{\rmr, \frac{1}{\lambda_{12}}}  ( \tilR_1,\tilR_2 ) .
\end{equation}
By using weak duality in   optimization theory, it can be seen that
\begin{equation}
E_{\rmr, \frac{1}{\lambda_{12}}}  ( \tilR_1,\tilR_2  ) \le E_{\rms\rmp}(\tilR_{12}).
\end{equation}
Hence, contribution from the joint types in $\calA$ given by the calculation in \eqref{eqn:sanov} do not dominate. As a result, the  error exponent for decoding message 1 is dominated by the joint types in $\calA^c$, and so the exponential behaviors of the upper bounds of $e_1$ and $e_{1,\calA^c}$ are the same. We thus conclude that 
\begin{equation}
\liminf_{n\to\infty} \frac{ 1}{n}\log \frac{1}{e_1 }  \ge E_{\rmr, \frac{1}{\lambda_{12}}}^{-} ( \tilR_1,\tilR_2,R_2).
\end{equation}

\subsection{Undetected Error Probability for Message 2 at Terminal~$\calY$} \label{sec:und2}
An undetected error for message 2 occurs in one of two ways: (i) Step 1 succeeds but the declared message
pair $(\hatm_1, \hatm_2)$ is such that the second component is some natural number not equal to 1; (ii) Step 1 fails (which
we denote as event $\calF$) and Step 2 succeeds but the declared message in this step $\hatm_2$ is some natural number not
equal to 1. Thus, 
\begin{align}
\hate_2  &=\Pr\bigg( \bigcup_{\tilm_1\in\calM_1\tilm_2 \in\calM_2\setminus\{1\} }\calE(\tilm_1,\tilm_2)  \nn\\*
&\qquad  \qquad\qquad\qquad  \cup \Big(\calF \cap\bigcup_{\tilm_2 \in\calM_2\setminus\{1\}}\calG(\tilm_2) \Big)\bigg)
\end{align}
where $\calE(\tilm_1,\tilm_2)$ is defined in \eqref{eqn:Em1m2} and  
\begin{align}
\calG(\tilm_2)&:=\big\{ \hatI(U^n(\tilm_2)\wedge Y^n) \nn\\*
&\qquad \ge \tilR_2 +\lambda_2 \big|\hatI(U^n(1)\wedge Y^n) \ge R_2 \big|^+\big\}.
\end{align}

For Case (i), by using a similar calculation to that in Sec.~\ref{sec:und1}, we have   
\begin{align}
&\liminf_{n\to\infty}\frac{1}{n}\log\frac{1}{\Pr\big( \bigcup_{\tilm_1\in\calM_1\tilm_2 \in\calM_2\setminus\{1\} }\calE(\tilm_1,\tilm_2) \big)}  \nn\\*
&\ge E_{\rmr, \lambda_{12}}(R_1,R_2)+ \Delta_{12}.
\end{align}
An important point to note here is that the term $J_V(R_2)$ is {\em absent} because here we do not have to bound the probability  that the cloud center is decoded  correctly but the satellite codeword is decoded incorrectly.   So the exponent here is the {\em unpenalized} random coding error exponent for the ABC.

Now for Case (ii), we first analyze the probability that Step 1 fails, i.e., there is no unique $(\hatm_1, \hatm_2)$ satisfying \eqref{eqn:Step1}. This exponent is exactly that calculated in Sec.~\ref{sec:total1}. Thus,
\begin{align}
 \liminf_{n\to\infty}\frac{1}{n}\log\frac{1}{\Pr(\calF ) }\ge E_{\rmr, \frac{1}{\lambda_{12}}}^{-} ( \tilR_1,\tilR_2,R_2). \label{eqn:fail_exp}
\end{align}
Note that the {\em penalization is} present here because $(\tilm_1,\tilm_2)\ne (1,1)$ means there are three cases: (a)  $\tilm_1\ne 1,\tilm_2\ne 1$; (b)  $\tilm_1 = 1,\tilm_2\ne 1$; and (c)  $\tilm_1\ne 1,\tilm_2 = 1$. Finally, we need to bound the probability that the declared message in Step 2 is some natural number not equal to 1. This yields 
\begin{align}
 \liminf_{n\to\infty}\frac{1}{n}\log\frac{1}{ \Pr\big(\bigcup_{\tilm_2 \in\calM_2\setminus\{1\}}\calG(\tilm_2) \big) } \ge E_{\rmr,\lambda_2}(R_2)+\Delta_2. \vspace{-.03in} \label{eqn:undetected_2} 
 \end{align}
 So the exponent for Case (ii) is the maximum of the exponents derived in \eqref{eqn:fail_exp} and \eqref{eqn:undetected_2}.  Combining all these exponents yields the undetected error probability for message 2 in \eqref{eqn:e2hat_res}.

\subsection{Total Error Probability for Message 2 at Terminal~$\calY$}  \label{sec:total2}
Finally, we compute the total error probability for message 2. An error occurs  if and only if one of two events occurs: (i) Step 1 succeeds but message 2 is declared to be some $\hatm_2 \in\calM_1\setminus\{1\}$ (i.e., undetected error) or (ii) Step 1 fails and an error (undetected or erasure) occurs in Step~2.

For Case (i), the exponent of the error probability is $E_{\rmr , \lambda_{12}}(\tilR_1,\tilR_2 ) + \Delta_{12}$ without penalization because the cloud center $\tilm_2$   suffers from an undetected error. 

In Case (ii), Step 1 fails (event $\calF$ in Sec.~\ref{sec:und2}) and an error occurs in Step 2. Step 1 failing results in an error exponent of $ E_{\rmr, \frac{1}{\lambda_{12}}}^{-} ( \tilR_1,\tilR_2,R_2)$; cf.~\eqref{eqn:fail_exp}.  An error occurs in Step 2 with exponent $E_{\rmr, \frac{1}{\lambda_2}}(\tilR_2)$ by the same reasoning as the calculations in Sec.~\ref{sec:total1}. Combining these exponents yields~\eqref{eqn:e2_res}.
\vspace{-.05in}
\subsection*{Acknowledgements}
Discussions with Pierre Moulin and Silas L.\ Fong are gratefully acknowledged.   The author's research is supported by an NUS Young Investigator Award R-263-000-B37-133.
\vspace{-.05in}
\bibliographystyle{IEEEtran}
\bibliography{isitbib}
\end{document}